\documentclass{emulateapj}

\slugcomment{Accepted for Publication in ApJ Letters}

\shorttitle{[CII] in MIPS J1428}
\shortauthors{Hailey-Dunsheath et al.}

\begin{document}


\title{Detection of the $158$ $\mu$\lowercase{m} [CII] Transition at $\lowercase{z}=1.3$: Evidence for a Galaxy-Wide Starburst}





\author{S. Hailey-Dunsheath\altaffilmark{1,2}, T. Nikola\altaffilmark{1}, G. J. Stacey\altaffilmark{1}, T. E. Oberst\altaffilmark{1,3}, S. C. Parshley\altaffilmark{1}, D. J. Benford\altaffilmark{4}, J. G. Staguhn\altaffilmark{4,5,6}, and C. E. Tucker\altaffilmark{7}}


\altaffiltext{1}{Department of Astronomy, Cornell University, Ithaca, NY 14853.}
\altaffiltext{2}{Current Address: Max-Planck-Institut f{\"u}r extraterrestrische Physik, Postfach 1312, D-85741 Garching, Germany; steve@mpe.mpg.de.}
\altaffiltext{3}{Current Address: Department of Physics, Westminster College, 319 S. Market St., New Wilmington, PA 16172.}
\altaffiltext{4}{Observational Cosmology Laboratory (Code 665), NASA Goddard Space Flight Center, Greenbelt, MD 20771.}
\altaffiltext{5}{Department of Astronomy, University of Maryland, College Park, MD 20742.}
\altaffiltext{6}{Current Address: Department of Physics \& Astronomy, Johns Hopkins University, Baltimore, MD 21218.}
\altaffiltext{7}{School of Physics and Astronomy, Cardiff University, Cardiff CF24 3AA, UK.}


\begin{abstract}
We report the detection of $158$ $\mu$m [CII] fine-structure line emission from MIPS J142824.0+352619, a hyperluminous ($L_\mathrm{IR}\sim10^{13}$ $L_\sun$) starburst galaxy at $z=1.3$. The line is bright, and corresponds to a fraction $L_\mathrm{[CII]}/L_\mathrm{FIR}\approx2\times10^{-3}$ of the far-IR (FIR) continuum. The [CII], CO, and FIR continuum emission may be modeled as arising from photodissociation regions (PDRs) that have a characteristic gas density of $n\sim10^{4.2}$ cm$^{-3}$, and that are illuminated by a far-UV radiation field $\sim$$10^{3.2}$ times more intense than the local interstellar radiation field. The mass in these PDRs accounts for approximately half of the molecular gas mass in this galaxy. The $L_\mathrm{[CII]}/L_\mathrm{FIR}$ ratio is higher than observed in local ULIRGs or in the few high-redshift QSOs detected in [CII], but the $L_\mathrm{[CII]}/L_\mathrm{FIR}$ and $L_\mathrm{CO}/L_\mathrm{FIR}$ ratios are similar to the values seen in nearby starburst galaxies. This suggests that MIPS J142824.0+352619 is a scaled-up version of a starburst nucleus, with the burst extended over several kiloparsecs.
\end{abstract}





\section{Introduction}

The $^2$P$_{3/2}$ $\rightarrow$ $^2$P$_{1/2}$ transition of C$^+$ ($\lambda=157.74$ $\mu$m) is one of the brightest emission lines in star-forming galaxies, typically accounting for $0.1-1\%$ of the far-IR (FIR) continuum~\citep{Stacey1991,Malhotra2001}. Much of this emission arises from the warm and dense photodissociation regions (PDRs) that form on the UV-illuminated surfaces of molecular clouds. The [CII] transition is a primary PDR coolant, and is a sensitive probe of both the physical conditions of the photodissociated gas, and the intensity of the ambient stellar radiation field~\citep{Hollenbach1999}. In ultraluminous infrared galaxies (ULIRGs), the $L_\mathrm{[CII]}/L_\mathrm{FIR}$ ratio is a factor of $\sim$$7$ times lower than in less luminous systems~\citep{Luhman2003}, for reasons that are not well understood.

Large aperture (sub)millimeter telescopes have been used to search for [CII] emission from sources at $z>3$, and 3 FIR-luminous QSOs at $z=4.4-6.4$ have been detected thus far. SDSS J1148~\citep{Maiolino2005} and BR 1202N~\citep{Iono2006quasar} have low $L_\mathrm{[CII]}/L_\mathrm{FIR}$ ratios similar to or smaller than the mean local ULIRG value, while BRI 0952~\citep{Maiolino2009} has a somewhat larger ratio falling at the lower end of the range seen in normal galaxies. We have initiated a search for [CII] emission from galaxies at $z=1-2$, concentrating on star-formation--dominated systems selected by their FIR or submillimeter continuum brightness. Here we report our first detection, from the $z=1.3$ hyperluminous starburst galaxy MIPS J142824.0+352619 (hereafter MIPS J1428).

MIPS J1428 was identified in a Spitzer/MIPS blank field survey as a bright 160 $\mu$m source with red optical/near-IR colors, and subsequent spectroscopic and photometric measurements established it as a hyperluminous ($L_\mathrm{IR}[8-1000$ $\mu$m$]=3.2\times10^{13}$ $L_\sun$) galaxy at $z=1.325$~\citep{Borys2006}. There are several indications that this IR emission is powered by star formation, including the brightness of the PAH features and the nondetection in hard X-ray emission, and with a starburst origin the estimated $L_\mathrm{IR}$ corresponds to a star formation rate of $\sim$$5500$ $M_\sun$ yr$^{-1}$~\citep{Borys2006,Desai2006}. MIPS J1428 was detected in CO($2$$\rightarrow$$1$) and CO($3$$\rightarrow$$2$), and the line luminosities correspond to a large gas mass of $M_\mathrm{H_2}\sim10^{11}$ $M_{\sun}$~\citep{Iono2006CO}. However, a gravitational lens may amplify the flux by as much as a factor of 10~\citep{Borys2006,Iono2006SMA,Iono2006CO}, making the actual star formation rate and gas mass correspondingly lower. Independent of the unknown lensing amplification, the large value of $L_\mathrm{IR}/L\arcmin_\mathrm{CO}\approx320$ $L_\sun$ (K km s$^{-1}$)$^{-1}$~\citep{Iono2006CO} indicates that MIPS J1428 is forming stars with high efficiency, similar to that seen in local ULIRGs and high-redshift submillimeter galaxies (SMGs)~\citep{Solomon2005}.

This is the first detection of [CII] in the $z=1-3$ epoch of peak star formation activity in the Universe, and the first detection from a high-redshift galaxy not associated with a QSO.

\section{Observations}

We observed the [CII] line toward MIPS J1428 in March 2008 with ZEUS~\citep{Stacey2007,Hailey_Dunsheath2009thesis} at the Caltech Submillimeter Observatory (CSO)\footnote{The Caltech Submillimeter Observatory is supported by NSF contract AST-0229008} on Mauna Kea, Hawaii. ZEUS is a dual-band (350 and 450 $\mu$m) grating spectrometer equipped with a $1\times32$ semiconductor bolometer array. For this observing run the instrument was configured as described in~\citet{Hailey-Dunsheath2008}. At the redshifted wavelength ($\lambda=366.75$ $\mu$m) of the [CII] transition the spectrometer provides a slit-limited resolving power of $\lambda/\Delta\lambda\approx900$, and the 14 detectors operating in the 350 $\mu$m window combine to provide an instantaneous bandwidth corresponding to $\delta v=2825$ km s$^{-1}$.

\begin{figure}
\epsscale{1.1}
\plotone{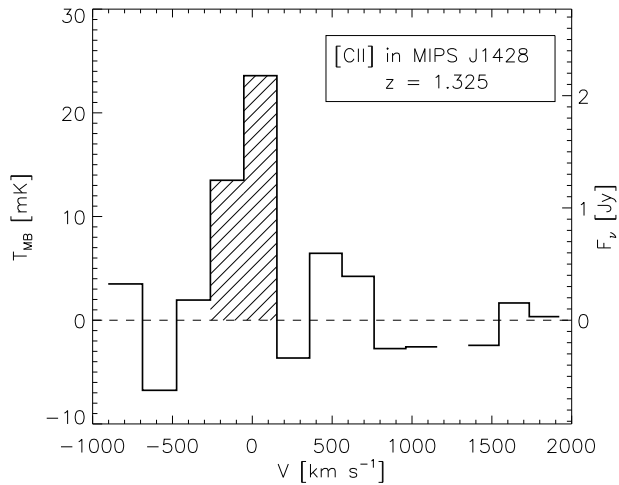}
\caption{ZEUS/CSO spectrum of the $158$ $\mu$m $\mathrm{[CII]}$ transition in MIPS J1428. The velocity scale is referenced to $z=1.325$, and the pixel centered at $v=1250$ km s$^{-1}$ is removed due to the presence of a telluric absorption feature. The integrated intensity is obtained by summing the 2 central spectral bins, which cover $\Delta v=416$ km s$^{-1}$. The line is detected at 6$\sigma$ significance. \label{fig1}}
\end{figure}

Orion (BN-KL) was used as spectral calibrator, and Mars~\citep[$T_B=213$ K;][]{Hildebrand1985} as flux calibrator. We integrated on MIPS J1428 for a total of $135$ minutes (including chopping) in good submillimeter weather, with $\tau_{350\,}$$_\mathrm{\mu m}=1.0-1.5$. The $13\arcsec$ ZEUS/CSO beam was centered at $\mathrm{R.A.}=14^\mathrm{h}28^\mathrm{m}24\fs1$, $\mathrm{decl.}=+35\arcdeg26\arcmin19\arcsec$ (J2000.0), and the data were obtained by chopping and nodding the telescope with a 30$\arcsec$ throw. A linear baseline is removed from the final spectrum, shown in Figure~\ref{fig1}. The integrated line flux is $F_\nu\Delta v=713$ Jy km s$^{-1}$, and we compare this with other observations in Table~\ref{mips_tracers}. With $L_\mathrm{FIR}=2.6\times10^{13}$ $L_\sun$~\citep{Borys2006}, we estimate $L_\mathrm{[CII]}/L_\mathrm{FIR}\approx2.1\times10^{-3}$. 

\section{Results} \label{resultssection}
\subsection{Origins of the [CII] Line} \label{ciiorigins}

Carbon has a lower ionization potential (11.26 eV) than hydrogen, and consequently the [CII] transition is an important coolant of both ionized gas and diffuse atomic gas. However, most of the [CII] emission from nearby IR-bright galaxies arises from dense PDRs illuminated by the far-UV (FUV) radiation from young stars~\citep{Crawford1985,Stacey1991}. Combined HII region and PDR modeling of the starburst templates NGC 253~\citep{Carral1994} and M82~\citep{Lord1996,Colbert1999} has demonstrated that PDRs account for at least 70\% of the [CII] emission in these sources. The available data suggests that such starburst galaxies are the best local analogs of MIPS J1428 (see section~\ref{discussion}), so here we assume that 70\% of the [CII] emission from MIPS J1428 arises in PDRs. We further assume that PDRs produce all of the CO($2$$\rightarrow$$1$), CO($3$$\rightarrow$$2$), and FIR continuum emission in this source, and we model these tracers as arising from a single representative PDR. 

\begin{table}
\begin{center}
\caption{Spectral Line and Continuum Observations of MIPS J1428\label{mips_tracers}}
\begin{tabular}{cccccrrrrr}
\tableline\tableline
Tracer	                  & Flux       					& $L$				& $\sigma$ & ref  \\    
                          & [Jy km s$^{-1}$]				& [$L_{\sun}$]			& [\%] &      \\
\tableline
$\mathrm{[CII]}$	  & 713          				& $5.4$ $\times$ $10^{10}$	& 30 & 1    \\
CO($2$$\rightarrow$$1$)   & 5.3						& $4.9$ $\times$ $10^{7}$	& 23 & 2    \\
CO($3$$\rightarrow$$2$)   & 13.9					& $1.9$ $\times$ $10^{8}$	& 35 & 2    \\
IR (8-1000 $\mu$m)        & ...						& $3.2$ $\times$ $10^{13}$	& 22 & 3    \\
FIR\tablenotemark{a}	  & ...						& $2.6$ $\times$ $10^{13}$	& 22 & 3    \\
\tableline
\end{tabular}
\tablecomments{$\mathrm{L}$uminosities are uncorrected for lensing, and are calculated using $\Omega_M=0.27$, $\Omega_\Lambda=0.73$, $H_0=71$ km s$^{-1}$ Mpc$^{-1}$. References: (1) this work, (2)~\citet{Iono2006CO}, (3)~\citet{Borys2006}.}
\tablenotetext{a}{We estimate the rest-frame 60 $\mu$m and 100 $\mu$m luminosity densities from the~\citet{Borys2006} model SED, and calculate $L_\mathrm{FIR}$ following $F_\mathrm{FIR}=1.26\times10^{-14}$ $(2.58f_\mathrm{60}+f_\mathrm{100})$ W m$^{-2}$~\citep{Sanders1996}.}
\end{center}
\end{table}

\subsection{PDR Analysis} \label{pdranalysis}

We use the PDR models of~\citet{Kaufman1999}, in which the emission is determined by the gas density, $n$, and the incident FUV (6 eV $<h\nu<$ 13.6 eV) flux, $G_0$ (expressed in units of the Habing Field, $1.6\times10^{-3}$ ergs cm$^{-2}$ s$^{-1}$). These models calculate the line emission generated by a cloud illuminated on only one side, and must be modified to model clouds in the active regions of galaxies that are illuminated on all sides. For this geometry,~\citet{Kaufman1999} note that an observer will detect optically thin radiation emitted by both the near and far sides of the cloud, while optically thick emission will only be visible from the near side. The [CII] line is generally moderately optically thin in these calculations, and the FIR continuum is also assumed to be optically thin, but the low-$J$ CO transitions are optically thick for the nominal $A_V=10$ cloud depth. We therefore increase the measured CO fluxes by a factor of 2 with respect to these other tracers before comparing with the models. 

\begin{figure}
\epsscale{1.1}
\plotone{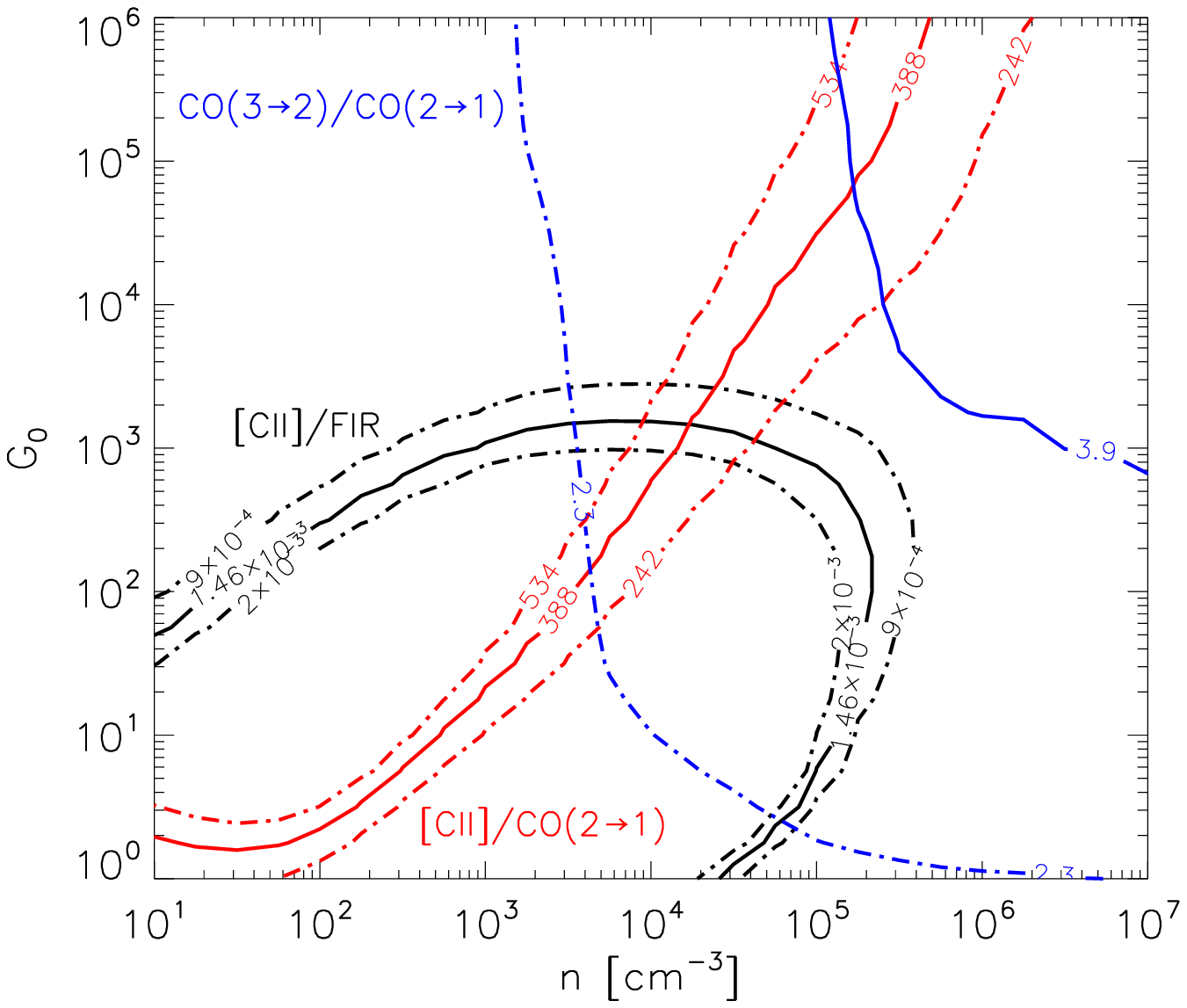}
\caption{$L_\mathrm{[CII]}/L_\mathrm{FIR}$, $L_\mathrm{[CII]}/L_{\mathrm{CO}(2\rightarrow1)}$, and $L_{\mathrm{CO}(3\rightarrow2)}/L_{\mathrm{CO}(2\rightarrow1)}$ as a function of density ($n$) and FUV flux ($G_0$), from the PDR models of~\citet{Kaufman1999}. Contours are drawn for the measured values (\textit{solid}) and $\pm1\sigma$ uncertainties (\textit{dot-dashed}) of MIPS J1428, after scaling the [CII] flux by 70\% to account for a non-PDR contribution, and increasing the CO fluxes by a factor of 2 to account for geometrical effects (see section~\ref{resultssection}).\label{fig2}}
\end{figure}

In Figure~\ref{fig2} we show the values of $n$ and $G_0$ allowed by the observations listed in Table~\ref{mips_tracers}, after correcting the [CII] and CO measurements as described above. In the $n\approx10^{3}-10^{5}$ cm$^{-3}$ range the $L_\mathrm{[CII]}/L_\mathrm{FIR}$ ratio provides a strong constraint on $G_0$, while the $L_\mathrm{[CII]}/L_{\mathrm{CO}(2\rightarrow1)}$ ratio is primarily a function of density. The best fit solution obtained from these two ratios is consistent with the $n\gtrsim10^{3.5}$ cm$^{-3}$ lower limit imposed by the $L_{\mathrm{CO}(3\rightarrow2)}/L_{\mathrm{CO}(2\rightarrow1)}$ ratio, and we conclude that the relative strengths of the [CII], CO($2$$\rightarrow$$1$), CO($3$$\rightarrow$$2$), and FIR continuum emission may be reproduced by a PDR with $n\sim10^{4.2}$ cm$^{-3}$ and $G_0\sim10^{3.2}$. Reducing the PDR contribution to the total [CII] emission by a factor of 2 (3) would increase $n$ by a factor of 7.2 (22), but $G_0$ by only a factor of 1.6 (1.8). We also note that without the factor of 2 correction to the observed CO fluxes, $n$ would be 4.2 times lower, but $G_0$ would remain unchanged.

An estimate of the atomic gas mass associated with the PDRs in MIPS J1428 is obtained from the inferred [CII] luminosity and this PDR solution. Assuming the [CII] emission is optically thin, this mass is given as
\begin{eqnarray}
\frac{M_a}{M_{\sun}}=&&0.77 \biggl( \frac{0.7\,L_\mathrm{[CII]}}{L_{\sun}} \biggr) \biggl( \frac{1.4 \times 10^{-4}}{X_\mathrm{C^+}} \biggr) \nonumber \\
&&\times \biggl( \frac{1 +2\, \mathrm{exp}(-91\,\mathrm{K}/T) + n_\mathrm{crit}/n}{2\, \mathrm{exp}(-91\,\mathrm{K}/T)} \biggr),
\end{eqnarray}
\noindent where we use a C$^+$ abundance per hydrogen atom of $X_\mathrm{C^+}=1.4\times10^{-4}$~\citep{Savage1996}. The surface temperature of a PDR with $n=10^{4.2}$ cm$^{-3}$ and $G_0=10^{3.2}$ is $T_S\approx230$ K~\citep{Kaufman1999}. Assuming this temperature is representative of the full C$^+$ region, and adopting a critical density of $n_\mathrm{crit}=2.7\times10^3$ cm$^{-3}$~\citep{Launay1977}, this expression yields $M_a=5.5\times10^{10}$$\mu^{-1}$ $M_{\sun}$ (where $\mu$ is the lensing amplification). This is 55\% of the molecular gas mass estimated from the CO luminosity~\citep{Iono2006CO}, which is similar to the PDR mass fractions in starburst galaxies~\citep[][corrected for the smaller $X_\mathrm{C^+}$ used here]{Stacey1991}. Much like in nearby starburst nuclei, the PDR mass in MIPS J1428 is a significant fraction of the total molecular gas mass.

\section{Discussion} \label{discussion}
\subsection{[CII], CO, and the FIR Continuum: A Comparison to Other Galaxies} \label{starburst}

\begin{figure*}
\epsscale{1.1}
\plotone{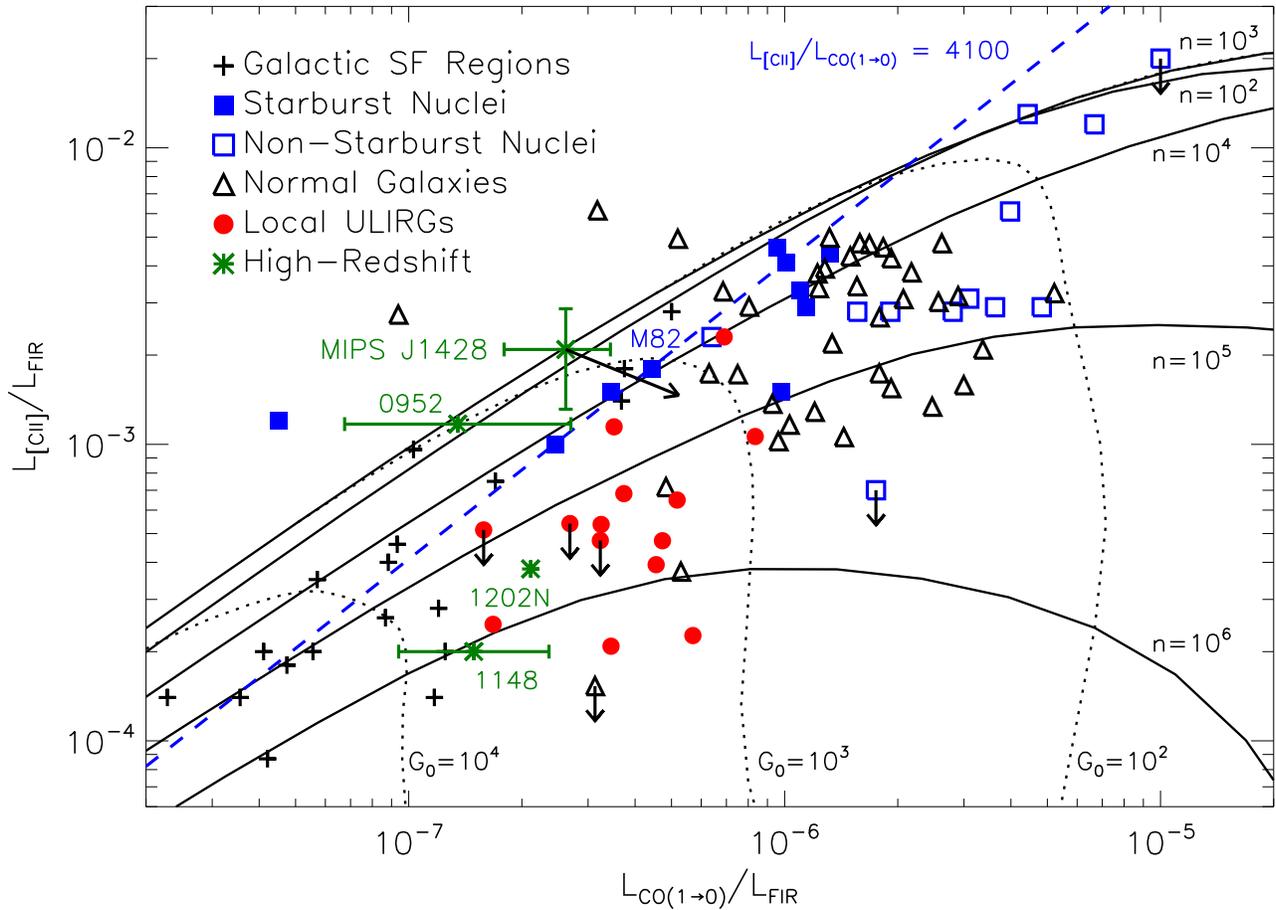}
\caption{$L_\mathrm{[CII]}/L_\mathrm{FIR}$ vs. $L_{\mathrm{CO}(1\rightarrow0)}/L_\mathrm{FIR}$ for Galactic star-forming regions (\textit{crosses}), starburst nuclei (\textit{filled squares}), non-starburst nuclei (\textit{open squares}), normal galaxies (\textit{triangles}), local ULIRGs (\textit{circles}), and high-redshift sources (\textit{asterisks}). CO($1$$\rightarrow$$0$) luminosities of the normal galaxies and local ULIRGs are taken from the literature and from J. Graci{\'a}-Carpio et al. (2010, in preparation). CO($1$$\rightarrow$$0$) luminosities for MIPS J1428 and 2 of the other high-redshift sources are estimated from the measurements of higher-$J$ lines as described in the text. Starburst galaxies and Galactic star-forming regions are characterized by $L_\mathrm{[CII]}/L_{\mathrm{CO}(1\rightarrow0)}\approx4100$ (\textit{dashed}). Overplotted are the PDR model calculations for gas density ($n$) and FUV field strength ($G_0$) from~\citet{Kaufman1999}, and a vector indicates how the MIPS J1428 data point was shifted for the PDR analysis in section~\ref{pdranalysis}.\label{fig3}}
\end{figure*}

To place our [CII] measurement of MIPS J1428 in context, we compare the [CII], CO, and FIR continuum emission from this source with that from other galaxies and Galactic sources. In Figure~\ref{fig3} we plot the $L_\mathrm{[CII]}/L_\mathrm{FIR}$ and $L_{\mathrm{CO}(1\rightarrow0)}/L_\mathrm{FIR}$ ratios for samples of Galactic star-forming regions and nearby gas-rich galaxies~\citep{Stacey1991}, normal galaxies~\citep{Malhotra2001}, local ULIRGs~\citep{Luhman2003}, and MIPS J1428 and the 3 other high-redshift sources detected in [CII]~\citep{Maiolino2005,Iono2006quasar,Maiolino2009}. We estimate the CO($1$$\rightarrow$$0$) luminosity of MIPS J1428 assuming an $L\arcmin_{\mathrm{CO}(2\rightarrow1)}/L\arcmin_{\mathrm{CO}(1\rightarrow0)}=0.9$ ratio intermediate between the ratios observed in local ULIRGs~\citep[$\approx$$0.7$;][]{Downes1998} and in starburst nuclei~\citep[$\approx$$1.1$;][]{Harrison1999,Weiss2005M82}. For reference, we overplot the PDR model curves from~\citet{Kaufman1999}. A vector attached to the MIPS J1428 data point indicates how the measured ratios were adjusted for the PDR analysis in section~\ref{pdranalysis} (to account for non-PDR contributions to the [CII] emission, and optical depth effects in the CO transititions), and we further note that MIPS J1428 is placed in Figure~\ref{fig3} assuming an $L\arcmin_{\mathrm{CO}(2\rightarrow1)}/L\arcmin_{\mathrm{CO}(1\rightarrow0)}$ ratio $\approx$$20\%$ lower than our best fit PDR solution. We compare MIPS J1428 with each of the populations shown in Figure~\ref{fig3} in turn.

\citet{Stacey1991} study the [CII] emission from the central regions of a sample of nearby galaxies, and find that the $L_\mathrm{[CII]}/L_{\mathrm{CO}(1\rightarrow0)}$ and $L_\mathrm{[CII]}/L_\mathrm{FIR}$ ratios are sensitive tracers of the star formation activity in these sources. Starburst galaxies are well characterized by the same $L_\mathrm{[CII]}/L_{\mathrm{CO}(1\rightarrow0)}=4100$ ratio seen in Galactic star-forming regions, which is a factor of $\sim$$3$ times larger than in non-starburst galaxies and Galactic molecular clouds (Fig.~\ref{fig3}). 
However, the $L_\mathrm{[CII]}/L_\mathrm{FIR}$ ratios in starburst galaxies are higher than those observed in the most intense Galactic OB star formation regions. These results imply that while much of the molecular gas in a starburst nucleus is photodissociated by the FUV radiation from young stars, the very intense fields found in the most extreme Galactic star-forming regions are not providing the bulk of the luminosity. This picture is supported by PDR modeling, which indicates that the FUV fields producing the emission in the centers of starburst galaxies are intermediate in intensity between those producing the emission in non-starburst galaxies, and those found in Galactic OB star formation regions (Fig.~\ref{fig3}). As the data point for MIPS J1428 falls near the high-$G_0$ end of the distribution of starburst points in Figure~\ref{fig3}, we conclude that in MIPS J1428, like in nearby starburst nuclei, the bulk of the molecular gas is exposed to moderately-intense FUV radiation produced by young stars. 

\citet{Malhotra2001} observe [CII] and other FIR fine-structure line emission from a sample of 60 normal galaxies with varying levels of star formation activity, which are all sufficiently distant that the FIR emission is contained within the $\approx$$70\arcsec$ \textit{ISO}-LWS beam. These authors find that the FIR continuum and much of the [CII] emission arises from PDRs, which in many of their galaxies are characterized by similarly elevated values of $G_0$ as deduced here for MIPS J1428. However, the~\citet{Malhotra2001} sources have lower $L_\mathrm{[CII]}/L_\mathrm{CO}$ ratios than MIPS J1428 or its starburst analogs, indicating that most of the CO emission is produced by molecular gas residing in less active star-forming regions than those that produce the fine-structure line emission. This difference helps reinforce our conclusion that in MIPS J1428 it is the entire galaxy, rather than just an active central region, that is host to vigorous star formation.

The median $L_\mathrm{[CII]}/L_\mathrm{FIR}$ ratio in local ULIRGs is a factor of $\sim$$7$ times lower than in normal star-forming galaxies~\citep{Luhman2003}, and a factor of $\sim$$4$ times lower than in MIPS J1428. One possible explanation for this low ratio is that the [CII] emission is produced in dense PDRs illuminated by intense FUV radiation~\citep[e.g.,][]{Papadopoulos2007}, as indicated by the PDR model overlays in Figure~\ref{fig3}. Alternatively, the [CII] emission may originate in PDRs with more modest values of $n$ and $G_0$, but an additional source of FIR continuum emission may lower the global $L_\mathrm{[CII]}/L_\mathrm{FIR}$ ratio~\citep{Luhman2003,Abel2009}.  Regardless of the origins of the weak [CII] emission in local ULIRGs, it is clear that these galaxies provide poor templates for MIPS J1428.

We also include the 3 high-redshift QSOs detected in [CII] in Figure~\ref{fig3}. SDSS J1148 has been detected in CO($3$$\rightarrow$$2$) and BRI 0952 in CO($5$$\rightarrow$$4$), and we use the measured $L\arcmin_{\mathrm{CO}(3\rightarrow2)}/L\arcmin_{\mathrm{CO}(1\rightarrow0)}=0.72$ and modeled $L\arcmin_{\mathrm{CO}(5\rightarrow4)}/L\arcmin_{\mathrm{CO}(1\rightarrow0)}=0.58$ ratios for Mrk 231~\citep{Papadopoulos2007} to estimate the CO($1$$\rightarrow$$0$) luminosities for these sources. To account for the uncertainties in the CO excitation of high-redshift systems~\citep[e.g.,][]{Hainline2006}, we show these data points with errors bars corresponding to 0.2 dex and 0.3 dex, respectively. The low $L_\mathrm{[CII]}/L_\mathrm{FIR}$ and $L_\mathrm{CO}/L_\mathrm{FIR}$ ratios in SDSS J1148 and BR 1202N indicate similar physical conditions as in local ULIRGs. Indeed,~\citet{Maiolino2005} found that the [CII], CO, and FIR continuum emission of SDSS J1148 is consistent with a high-density, high-$G_0$ PDR, and~\citet{Walter2009} showed that the star formation in this system is confined to a compact region with a high surface brightness comparable to that at the center of Arp 220. However, BRI 0952 has a larger $L_\mathrm{[CII]}/L_\mathrm{FIR}$ ratio and a $L_\mathrm{[CII]}/L_\mathrm{CO}$ ratio similar to that in MIPS J1428. We suggest that while SDSS J1148 and BR 1202N are likely dissimilar to MIPS J1428, the connection drawn here between MIPS J1428 and local starburst nuclei may also provide insight into the nature of the host galaxy of BRI 0952.

Finally, it is important to note that while we assume that the $L_\mathrm{[CII]}/L_\mathrm{FIR}$ and $L_\mathrm{CO}/L_\mathrm{FIR}$ ratios in MIPS J1428 are unaffected by the potential gravitational lensing, we cannot rule out the possibility that these ratios are altered by differential magnification. For example, one could construct a model in which the intrinsic emission ratios of MIPS J1428 are similar to those in local ULIRGs, but where the $L_\mathrm{[CII]}/L_\mathrm{FIR}$ ratio is enhanced by the relatively large magnification of a localized HII region that falls near a caustic. However, in the absence of other evidence supporting such a scenario, we assume that the measured flux ratios are equal to the intrinsic luminosity ratios.

\subsection{MIPS J1428 as an Extended Starburst}

The molecular gas in MIPS J1428 is exposed to similar UV radiation fields, and has similar excitation, as the gas in the central region of a typical starburst galaxy, and we suggest that MIPS J1428 may be modeled as a scaled-up version of a starburst nucleus. As an example, we consider scaling up the central region of the prototypical starburst galaxy M82 in such a manner as to match the much larger luminosity of MIPS J1428, while at the same time conserving the strength of the characteristic FUV radiation field that controls the PDR emission. We assume that in a starburst environment this characteristic field is determined by the global star formation density, rather than by the purely local properties of the interactions between individual star clusters and their natal molecular clouds. This interpretation was adopted for M82 and NGC 253 based on multiple-line PDR analysis~\citep{Wolfire1990,Carral1994,Lord1996}, and is also motivated by the non-linear relation between $I_\mathrm{[CII]}$ and $I_\mathrm{FIR}$ observed in larger samples of galactic nuclei~\citep{Crawford1985,Stacey1991}.

For a starburst region in which young stellar clusters and molecular clouds are randomly distributed,~\citet{Wolfire1990} modeled the relationship between the average FUV flux incident on the molecular gas ($G_0$), and the size ($D$) and total luminosity ($L_\mathrm{IR}$) of the region. This relationship depends on the mean free path of a FUV photon ($\lambda$), but in the limits of $\lambda\ll D$ and $\lambda\gg D$ these models give $G_0\propto L_\mathrm{IR}/D^3$ and $G_0\propto L_\mathrm{IR}/D^2$, respectively. Adopting $L_{IR}\sim3\times10^{10}$ $L_\sun$ for the central $D\sim300$ pc starburst region of M82~\citep{Telesco1980,Joy1987}, and assuming that MIPS J1428 and M82 have the same average $G_0$ (Fig.~\ref{fig3}), the factor of $\sim$$1000$$\mu^{-1}$ greater luminosity of MIPS J1428 then translates into a size scale of $D\approx3$$\mu^{-1/3}-10$$\mu^{-1/2}$ kpc. In short, if the large luminosity of MIPS J1428 is produced by only moderate-intensity radiation, the star formation generating this radiation must be extended over a large area.

While MIPS J1428 remains unresolved in dust continuum and molecular gas emission, integral field spectroscopy shows H$\alpha$ emission extended over $\approx$$0.7\arcsec$ ($\approx$$6$ kpc)~\citep{Swinbank2006}. This result is consistent with our interpretation of MIPS J1428 as a source powered by extended star formation, although the degree to which the H$\alpha$ image is distorted by the foreground lens, or by the presence of large and spatially-varying extinction, is unknown. Further interpretation of the connection between the bright [CII] emission in MIPS J1428 and the potentially large extent of the star formation will benefit from a size measurement in an extinction-free tracer, and better constraints on the lensing magnification.

If MIPS J1428 is indeed powered by a starburst extending over several kiloparsecs this would be in sharp contrast to local ULIRGs, in which most of the emission traced by CO interferometry is contained within the central $\sim$$1$ kpc~\citep{Downes1998}. However, there is ample evidence that the most luminous star-forming galaxies at higher redshifts are more extended. The best studied sample of such sources are the SMGs, which like MIPS J1428 are submillimeter-bright galaxies with luminosities of $L_\mathrm{IR}\sim10^{13}$ $L_\sun$ that are generated predominantly by star formation~\citep{Alexander2005ApJ}. Interferometric imaging of the radio continuum of $z=1-3$ SMGs has shown that many of these sources have detectable emission extended over $\sim$$10$ kpc~\citep{Chapman2004extended}, and that the population has a median FWHM size of $\sim$$5$ kpc~\citep{Biggs2008}. Similarly, many of the SMGs studied by~\citet{Tacconi2006,Tacconi2008} have CO FWHM sizes of $\sim$$4$ kpc. The model of extended star formation we suggest here for MIPS J1428 is therefore consistent with the large sizes of SMGs, which indicate that the most luminous star-forming galaxies at $z=1-3$ are powered by starbursts extending over much larger regions than in local systems.


\acknowledgments

We thank Javier Graci{\'a}-Carpio for kindly providing unpublished CO data, and the CSO staff for their support of ZEUS operations. We also thank an anonymous referee for many helpful comments on an earlier draft of this manuscript. This work was supported by NSF grants AST-0096881, AST-0352855, AST-0705256, and AST-0722220, and by NASA grants NGT5-50470 and NNG05GK70H.

\end{document}